\documentclass[letterpaper, 10 pt, conference]{ieeeconf}  

\IEEEoverridecommandlockouts                              

\usepackage{cite}
\usepackage{amsmath,amssymb,amsfonts}
\usepackage{algorithm}  
\usepackage{graphicx} 
\usepackage{subcaption}
\usepackage{graphicx}
\usepackage{textcomp}
\usepackage{xcolor}
\usepackage{mathrsfs}
\usepackage{amsmath}
\usepackage{algpseudocode}
\usepackage{graphicx}
\usepackage{hyperref}
\hypersetup{colorlinks,citecolor={blue}}
\usepackage{algorithmicx,algorithm}
\usepackage{algpseudocode}

\title{\LARGE \bf
A Simple Change Comparison Method for Image Sequences Based on Uncertainty Coefficient
}

\author{Ruzhang Zhao$^{1}$, Yajun Fang$^{2, *}$, Berthold K.P. Horn$^{2}$
\thanks{$^{1}$Ruzhang Zhao is with Department of Mathematical Science, Tsinghua University, 30 Shuangqing Rd.,Beijing, 100084, China.
        {\tt\small zrz6787@gmail.com}}%
\thanks{$^{2}$Yajun Fang and Berthold K.P. Horn are with the Computer Science and Artificial Intelligence Laboratory, Massachusetts Institute of Technology, 77 Massachusetts Avenue, Cambridge, Massachusetts 02139-4307, United States.
        {\tt\small yajufang, bkph@csail.mit.edu}}%
}

\begin{document}

\maketitle
\thispagestyle{empty}
\pagestyle{empty}

\begin{abstract}
For identification of change information in image sequences, most studies focus on change detection in one image sequence, while few studies have considered the change level comparison between two different image sequences. Moreover, most studies require the detection of image information in details, for example, object detection. Based on Uncertainty Coefficient(UC), this paper proposes an innovative method ``CCUC'' for change comparison between two image sequences. The proposed method is computationally efficient and simple to implement. The change comparison stems from video monitoring system. The limited number of provided screens and a large number of monitoring cameras require the videos or image sequences ordered by change level. We demonstrate this new method by applying it on two publicly available image sequences. The results are able to show the method can distinguish the different change level for sequences.
\par
{\it Key words}: Change Comparison, Change Detection, Image Sequence, Uncertainty Coefficient
\end{abstract}

\section{INTRODUCTION}
Change detection(CD) is a widely discussed topic in sequence analysis. Tremendous efforts have been devoted to detect the change in one sequence. A leading example focuses on the detection of abrupt change in a sequence of images with the same background under the order of time. The goal is to detect the set of pixels where there is difference between the last image and the previous images in the sequence. That is to say, we detect where the change happens. The corresponding tools for this kind of task have many applications, for example, remote sensing. Mainly driven by the need in remote sensing, many algorithms have been proposed since 1980's, such as \cite{singh1989review,coppin1996digital}.  Furthermore, applications of change detection include video surveillance \cite{lee2000introduction, stauffer2000learning, wren1997pfinder}, medical diagnosis and treatment \cite{bosc2003automatic,dumskyj1996accurate,lemieux1998detection,rey2002automatic,tacskesen2017unsupervised}, civil infrastructure \cite{landis1999technique,nagy2001volume}, under-water sensing \cite{edgington2003automated,lebart2000real,whorff1992video}, and driver assistance systems \cite{kan1996model,fang2003automatic}. In spite of different domains of application, the main task is to detect change in an image sequence with time order. There are many core algorithms which are ready to solve these tasks. Besides the methods applied in the above studies, we also referred to \cite{radke2005image,kasetkasem2002image,sivan1994change} for other methods in change detection. 
\par
In this paper, stemming from video monitoring system, we define and consider the change level comparison problem between two image sequences. In video monitoring system, people often deal with video surveillance, \cite{lee2000introduction, stauffer2000learning, wren1997pfinder} to detect change in videos. Here, we consider a daily task for video monitoring system, especially for system with limited resources. At most time, the video monitoring system with plenty of cameras is observed by limited number of screens, and the channels are changed from time to time randomly. However, the randomness may make the scenes with abrupt change ignored. The motivation of the paper is to provide a relative ranking for the change levels of different image sequences. Based on plenty of algorithms for change detection, it is easy to solve this problem with all the change detected, which may cost too much computation. And the efficiency is low with so many unnecessary details included. \par 
In \cite{viola1997alignment}, Viola and Wells III noted that using Mutual Information to detect the alignment between reference and target images required no exact position or posture detection. The larger the Mutual Information for the alignment is, the better the match is. This entropy based method largely reduces the computation for detecting pixel details of an image. Mutual Information is a measure of image matching that does not require the signals to be identical in the two images. Instead, it is a measure of how well the signal can be predicted in the second image, given the signal intensity in the first one. Invoked by \cite{viola1997alignment} and \cite{shannon2001mathematical}, the Mutual Information and Uncertainty Coefficient are used in this paper to formulate method for change comparison. Considering the multiple comparison can be simplified as binary comparison, this task is simplified as the change level comparison between two image sequences. In this study, we compare two images based on their pixel intensity. The change level is based on adjacent images. Then, we formulate the ``CCUC'' method, which means {\it Change Comparison method for image sequences based on Uncertainty Coefficient}. 
\par
The rest of the paper is organized as follows. Section 2 states the change comparison problem. The definition of change level and the solution applying Uncertainty Coefficient(UC) is formulated in Section 3. The paper is concluded and discussed in Section 4.

\section{PROBLEM FORMULATION}
\subsection{Problem Statement}
Assume there is an image sequence including N images that are taken at different times with the same scene. These images can contain any kind of scenes and any kind of activities. They are measured in pixel intensity. When considering two adjacent images in the same sequence, the spatial coordinates of the pixel locations in the two images may be different from each other. For example, object movement, texture defect, a slight shift of the background scene, or even the change of background brightness can result in different spatial coordinates. In this study, we assume the background is closely correlated while the object is independent of the change component. 
\par
Previous researches are interested in determining which position in the two correlated images is different to detect the change. In this study, we consider the requirement of video monitoring system. Thousands of monitoring cameras are available to determine whether there are emergent situations. However, only limited number of screens can be observed simultaneously. We would like to observe the specific monitoring camera that records more movement or other types of change. Therefore, priority or ranking should be offered to this monitoring system. If the solution to the change detection problem is applied to detect the exact objects, their edges, and their movement to analyze the possible change in the sequence, there is no doubt that it will work for change comparison. Here, we consider how to solve the problem without knowing the details of the images. For example, we do not know what kind of objects are in them.
In this study, we denote the N images in the sequence as a time series and each image as a random variable. Each random variable has a distribution. In this paper, we formulate the method ``CCUC"" to solve the change comparison problem.
\section{PROBLEM SOLUTION }
\subsection{Uncertainty Coefficient}
The entropy concept was firstly defined in \cite{shannon2001mathematical} in 1948, which was the fundament of information theory. To make things simple, the entropy related definitions are all formulated in discrete situation. The entropy of random variable $X$ is defined in Eq.~\ref{eq1}.
\begin{equation}
\operatorname {H}(X) = - \sum_{x\in\mathcal{X}} p(x)\log p(x),
\label{eq1}
\end{equation}
where $\mathcal{X}$ represents the set of values of random variable $X$. And the Mutual Information was used further in \cite{cronbach1955non}. The Mutual Information of two random variables is the sum of the entropy of one random variable and the conditional entropy of this one based on the other. It can also be defined as the sum of their entropies minus joint one. The joint entropy, conditional entropy and Mutual Information are formulated in Eq.~\ref{eq2}, Eq.~\ref{eq3} and Eq.~\ref{eq4} separately. 
\begin{equation}
\operatorname {H}(X,Y) = - \sum_{x\in\mathcal{X}, y\in\mathcal{Y}} p(x,y)\log p(x,y),
\label{eq2}
\end{equation}

\begin{equation}
\operatorname {H}(Y|X) = \sum_{x\in\mathcal{X}, y\in\mathcal{Y}} p(x,y)\log\frac{p(x)}{p(x,y)} =  \operatorname {H}(X,Y)-\operatorname {H}(X),
\label{eq3}
\end{equation}
where $\mathcal{X}$,  $\mathcal{Y}$ represent the set of values of random variables $X$ and $Y$ separately. 

\begin{equation}
\begin{aligned}
\operatorname {I} (X;Y) & =\sum _{y\in {\mathcal {Y}}}\sum _{x\in {\mathcal {X}}}{p(x,y)\log {\left({\frac {p(x,y)}{p(x)\,p(y)}}\right)}}\\
& = \operatorname {H}(X) + \operatorname {H}(Y) - \operatorname {H}(X,Y)\\
& = \operatorname {H}(X) - \operatorname {H}(X|Y).\\
\label{eq4}
\end{aligned}
\end{equation}

In order to use the relative value of Mutual Information, Uncertainty Coefficient(UC) is applied here. Uncertainty Coefficient(UC) was proposed based on the concept of entropy, in \cite{mills2011efficient,radke2005image,white2004performance}. The Uncertainty Coefficient is defined in Eq.~\ref{eq5}.
\begin{equation}
\operatorname {U}(X|Y) = \frac{\operatorname{I}(X;Y)}{\operatorname {H}(X)},
\label{eq5}
\end{equation}
where $\operatorname{I}(X;Y)$ and $\operatorname {H}(X)$ are defined in Eq.~\ref{eq4} and Eq.~\ref{eq1} separately.   
\subsection{Review of Mutual Information for Image Alignment}

Viola and Wells III used the Mutual Information for Image Alignment in \cite{viola1997alignment}. The main advantage of applying entropy based methods is the exact information in the image can be ignored. The image based on its pixel intensity is regarded as a kind of distribution. Thus, its entropy and Mutual Information is easy to calculate by Eq.~\ref{eq1} and Eq.~\ref{eq4} .
\par
In fact, Mutual Information is used for detecting the best alignment between a constructed model and real image data, or between a predicted image and real image data. The best projection is formulated as 
$\hat{T} = \arg\max_{T} \operatorname{I}(u(x);v(T(x)))$,
where u represents the model; the intensity of the model at point x is denoted as u(x); v represents the image; T represents the transformation from the model to the image.Transformation means a mapping from a specific point of the model to a coordinate point on the image. Thus, v(T(x)) denotes that, begin with x on the model, transform x to the image T(x), and then determine the intensity of T(x). By finding a transformation T from the model to image to maximize the Mutual Information between u(x) and v(T(x)), we determine the alignment between them.
\par
\begin{figure}
  \begin{subfigure}[b]{0.495\columnwidth}
    \includegraphics[width=\linewidth]{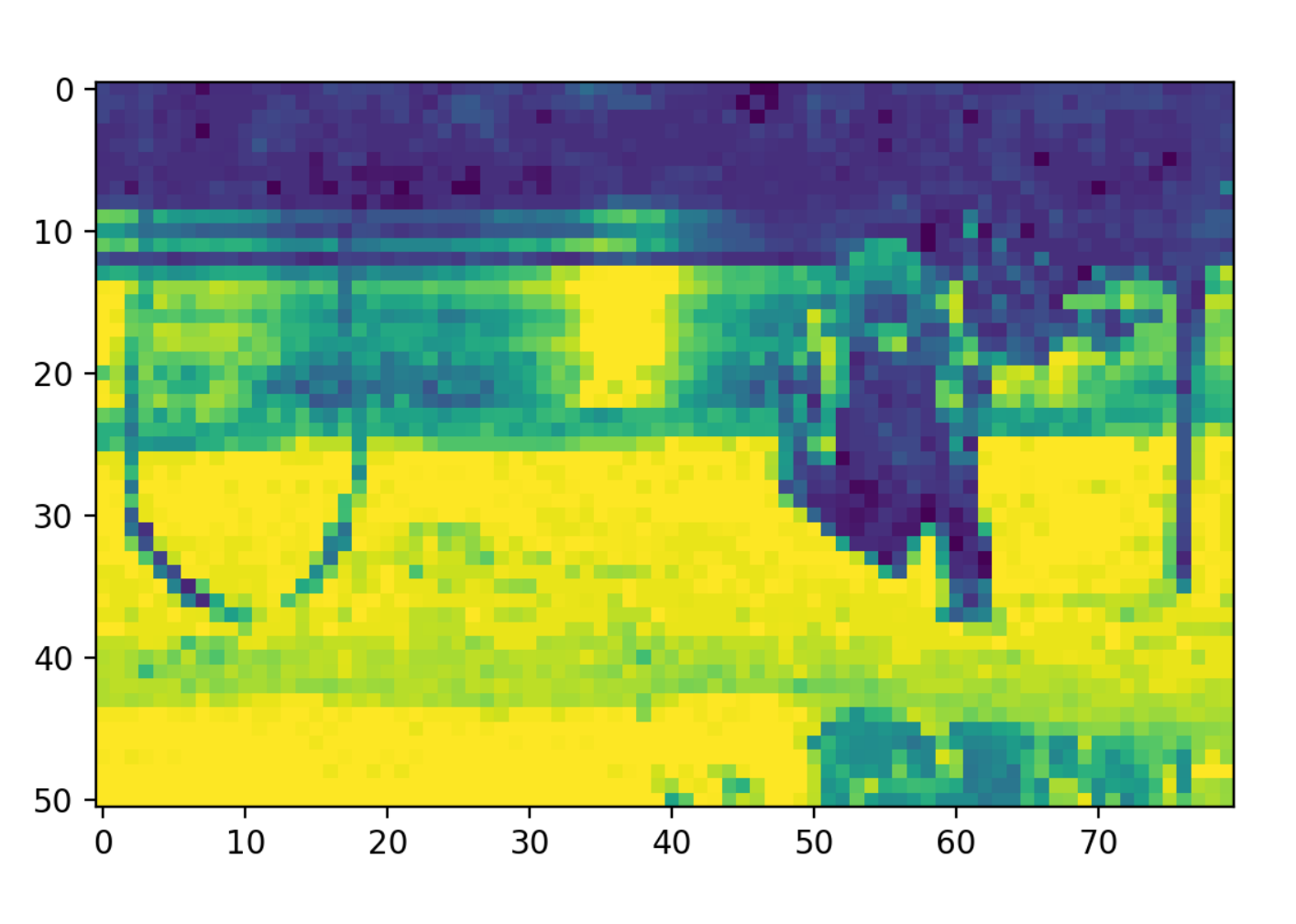}
    \caption{Image A}
    \label{fig:1}
  \end{subfigure}
  \hfill 
  \begin{subfigure}[b]{0.495\columnwidth}
    \includegraphics[width=\linewidth]{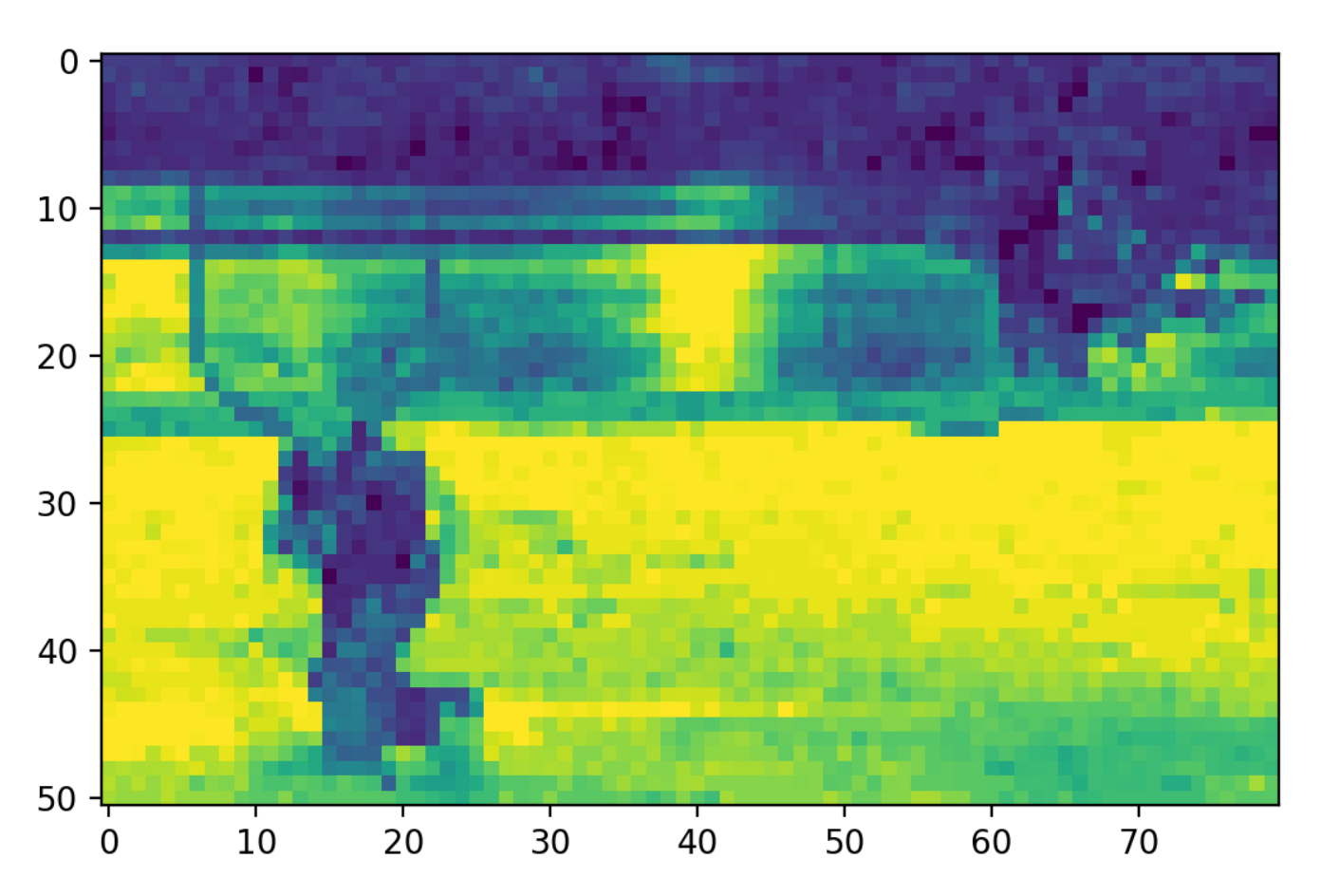}
    \caption{Image B}
    \label{fig:2}
  \end{subfigure}
  \caption{GIF1}
  \label{gp1}
\end{figure}

In \cite{penner2011sequence}, Penner asserted that the Mutual Information can be used not only in the alignment between the model and image but also between two images. We use the following two images in Fig.~\ref{gp1} to show how the Mutual Information represents the alignment between images. 
\par
Image A and B are from \href{https://i.gifer.com/9KxF.gif}{GIF1} and have the same background, which satisfies the basic assumption of the proposed problem. To reduce the computational effort, we reduce the size of the images to one fifth of the size of the original ones. Considering the image as a distribution based on the pixel intensity, we compute the Mutual Information between Image A and B. The Mutual Information between Image B and itself is also calculated, which is regarded as reference for perfect alignment. The results are listed in Table~\ref{tab0}.

\begin{table}[H]
\newcommand{\tabincell}[2]{\begin{tabular}{@{}#1@{}}#2\end{tabular}}
\caption{Mutual Information for Fig.~\ref{gp1}}
\begin{center}
\begin{tabular}{c|ccc}
\hline 
& Between A and B & Between B and B  \\
\hline
\tabincell{c}{Mutual\\ Information}& 11.9522 & 11.9883\\
\hline
\end{tabular}
\label{tab0}
\end{center}
\end{table}

The similarity  between the two Mutual Information results is attributed to the similar background of them. The results show that the alignment between A and B is worse than the alignment between B and itself, which matches the original setting. 

\subsection{Uncertainty Coefficient for Change Detection in One Image Sequence}

After dealing with two images, we focus on how to determine the most abrupt change in an image sequence. We apply the Uncertainty Coefficient(UC) between two images as a criterion for image alignment in a sequence. We need to notice the UC for two adjacent images is defined as: U(previous one $|$ later one). Based on the definition of UC in Eq.~\ref{eq5}, the Uncertainty Coefficient can be regraded as normalized entropy. In one sequence, considering the UCs between each two adjacent images, a series of UCs will be obtained. When using Mutual Information as a criterion for alignment, if the minimum point of the Mutual Information series is defined, it may be affected by the entropy of the two adjacent images, which is not very convincing for finding the point of abrupt change and not comparable among different sequences.  That is why we use Uncertainty Coefficient instead of Mutual Information for image sequences. 
\par
\begin{figure}[H]
\begin{center}
\includegraphics[width = 9cm]{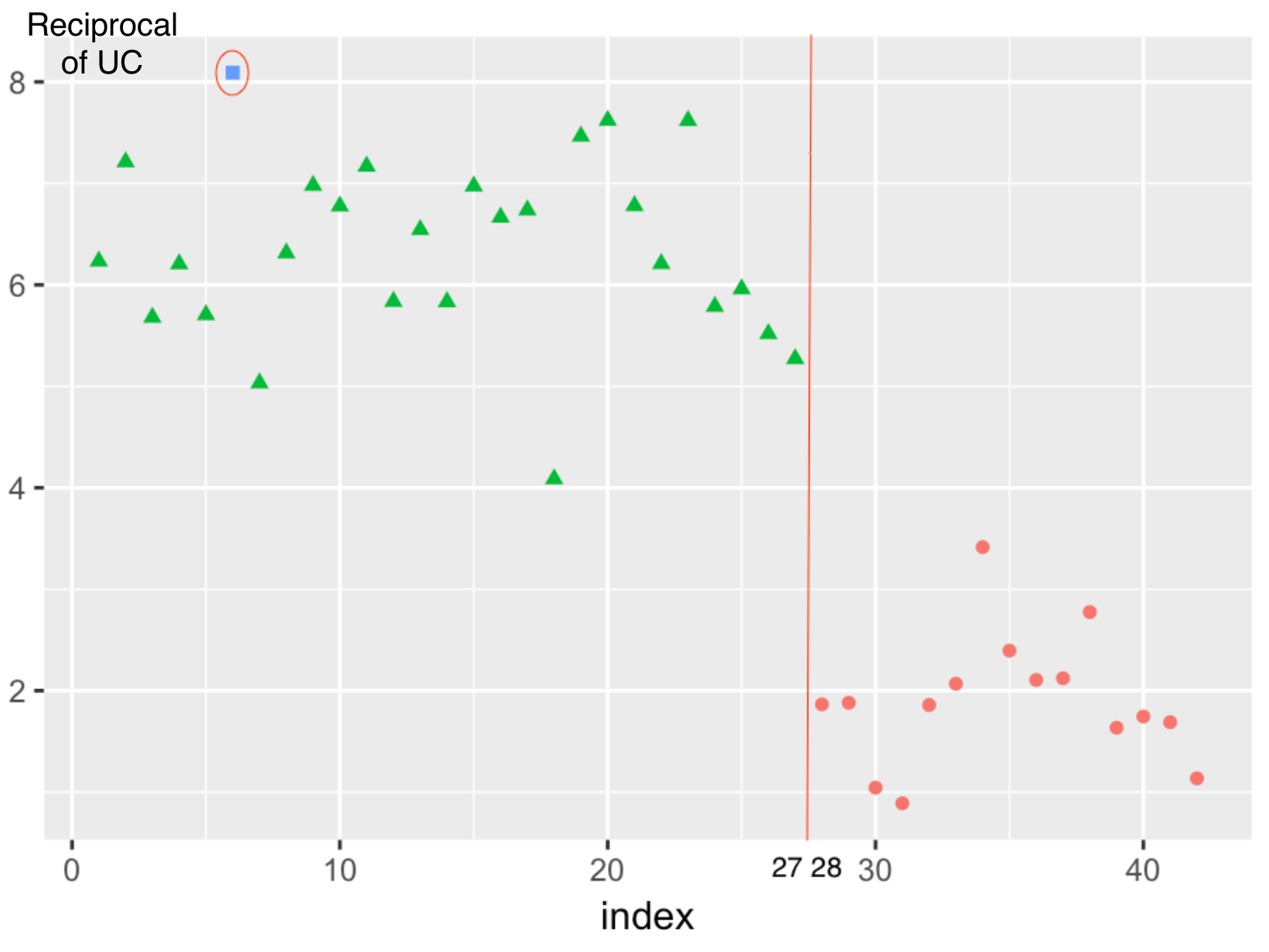}
\end{center}
\caption{Uncertainty Coefficient of GIF1}
\label{fig1}
\end{figure}

Thus, the definition of the most abrupt change point {\it Target} and the minimum of UCs {\it Target Value} in the image sequence are defined as follows:
\begin{equation}
\begin{aligned}
\operatorname{Target} & = \arg\min_{t,t+1 \in T}\operatorname{U}(J_{t}|J_{t+1}) \\
& = \arg\min_{t,t+1 \in T}\operatorname{I}(J_{t};J_{t+1})/\operatorname{H}(J_{t})\\
\end{aligned}
\label{target}
\end{equation}

where T represents the set of indexes for the image sequence with time series. We aim to obtain `t', `t+1', between which the most abrupt change occurs. We call the value $\min_{t,t+1 \in T}\operatorname{U}(J_{t}|J_{t+1})$ {\it Target Value of Sequence T} and refer to `t' as {\it Target}. \href{https://i.gifer.com/9KxF.gif}{GIF1} is used to show the results, and Image A and B are the first and last images of GIF1. The Uncertainty Coefficients of GIF1 are plotted in Fig.~\ref{fig1}.
where the y-axis is the reciprocal of UC. The larger the reciprocal is, the smaller the UC is. The highest point in Fig.~\ref{fig1} is the target we define in Eq.~\ref{target}. GIF1 shows a kid has a ride on a swing in the first half. In the second half, he accidentally drops himself to the ground. After we check, the red vertical line in Fig.~\ref{fig1} is exactly the boundary between swing activities and activities on the ground. It is clear to observe the movement on a swing is faster than the one on the ground. It is reasonable the UCs in the first half is smaller than the ones in the second half(The reciprocal of UC is the opposite). Thus, Uncertainty Coefficient is quite applicable for finding the most abrupt change in an image sequence.
\subsection{Uncertainty Coefficient for Comparison between Two Image Sequences}
\begin{figure}[H]
  \begin{subfigure}[b]{0.495\columnwidth}
    \includegraphics[width=\linewidth]{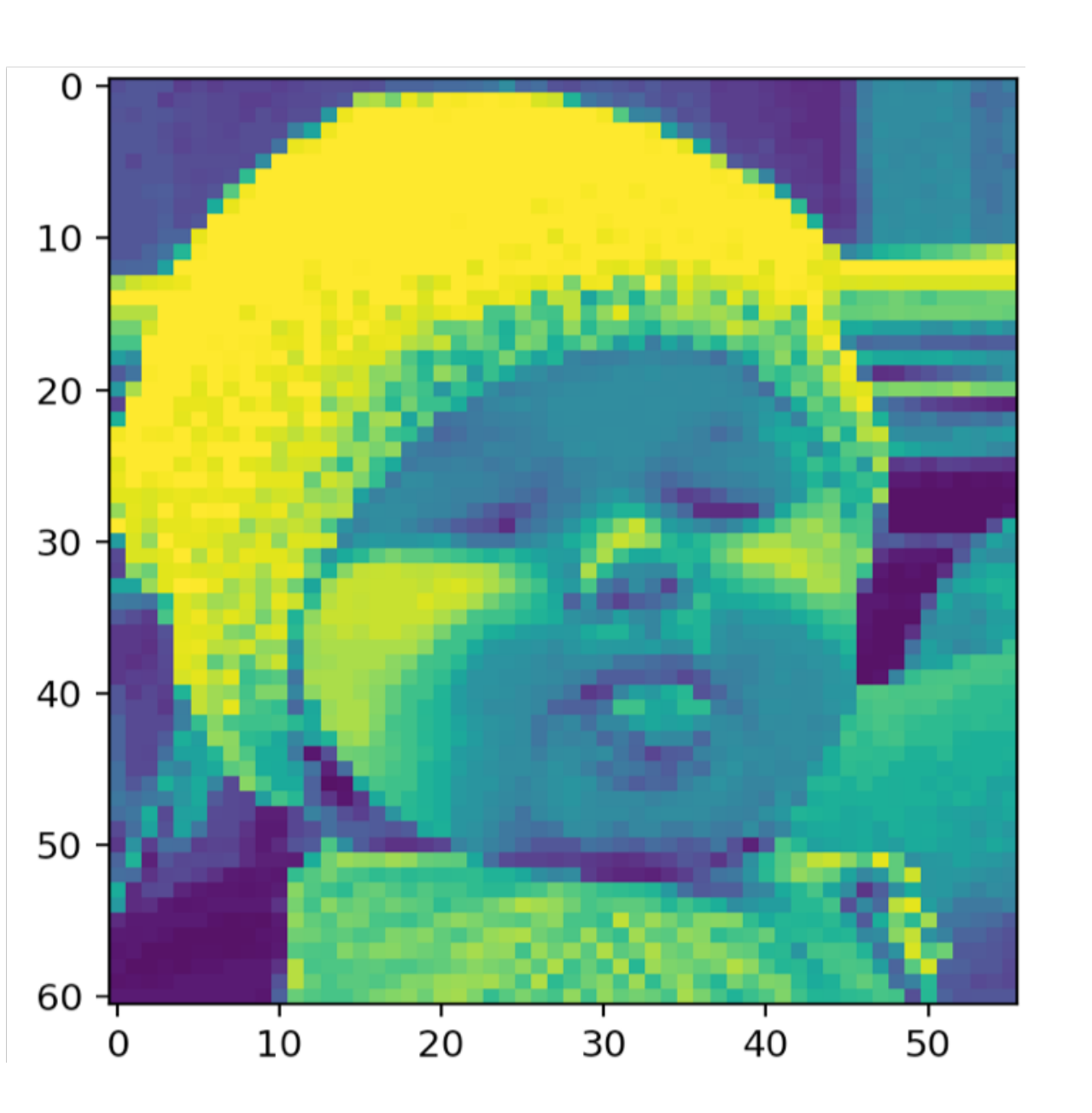}
    \caption{Image C}
    \label{fig:1}
  \end{subfigure}
  \hfill 
  \begin{subfigure}[b]{0.495\columnwidth}
    \includegraphics[width=\linewidth]{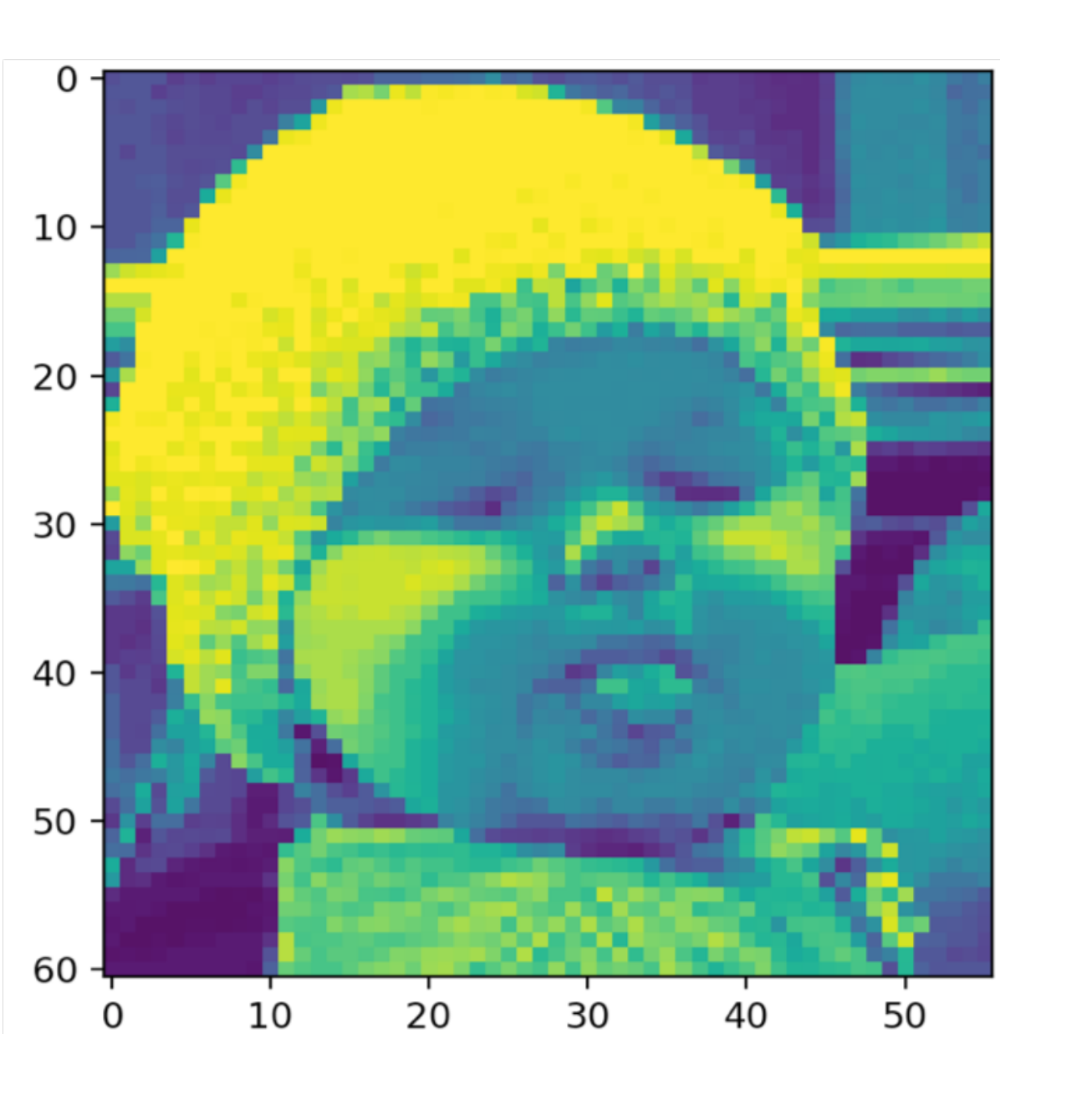}
    \caption{Image D}
    \label{fig:2}
  \end{subfigure}
  \caption{GIF2}
  \label{gif2}
\end{figure}

In this section, we focus on how to compare the change level between two sequences.
The \textbf{\underline{C}}hange \textbf{\underline{C}}omparison method for image sequences based on \textbf{\underline{U}}ncertainty \textbf{\underline{C}}oefficient (CCUC) is proposed.
 The Uncertainty Coefficient(UC) between two adjacent images is applied as a criterion for image alignment in an image sequence. We use the definition of target and target value in Eq.~\ref{target} to obtain the sequence with more abrupt change. By comparing the target value of two image sequences, the sequence with smaller target value is defined as the sequence with more abrupt change in Eq.~\ref{eq7}.
\par
\begin{equation}
\hat{T}=\arg\min_{T_{1},T_{2}}[\min_{t,t+1\in T}\operatorname{U}(J_{t}|J_{t+1})],
\label{eq7}
\end{equation}
\par
where $\hat{T}$ is the sequence with more abrupt change, $T_{1}$ and $T_{2}$ are two image sequences we used to compare. Intuitively, the Uncertainty Coefficient can show the relative change between two adjacent images compared with Mutual Information. Since UC is normalized entropy, UCs from different image sequences are comparable, for example, UC of the first two images in sequence A and UC of the last two images in sequence B. Thus, we use target value to obtain the comparison between two image sequences and  determine the one with  more abrupt change. Experiments are carried out and GIF1 and \href{http://5b0988e595225.cdn.sohucs.com/images/20171204/f80b90ef977e499b8e45e428aea7c06f.gif}{GIF2} are used. GIF2 includes only two images referred to as Images C and D in Fig.~\ref{gif2} and it is chosen to have almost no change compared with GIF1. Since there are only two images in GIF2, it is easy to obtain the target value. The reciprocals of target values for GIF1 and GIF2 are list in Table~\ref{tab1}.
In Table~\ref{tab1}, the target value of GIF1 is smaller than the one of GIF2, which is consistent with the GIFs we choose.

\begin{table}[H]
\newcommand{\tabincell}[2]{\begin{tabular}{@{}#1@{}}#2\end{tabular}}
\caption{Uncertainty Coefficient for GIF1 and GIF2}
\begin{center}
\begin{tabular}{c|ccc}
\hline 
&GIF1 & GIF2  \\
\hline
\tabincell{c}{Target Value \\ (reciprocal)}& 8.09& 3.16\\
\hline
\end{tabular}
\label{tab1}
\end{center}
\end{table}
Thus, based on the Target Values of different sequences, the sequence with more abrupt change can be obtained. The original purpose of our paper is to provide a reference for video monitoring system. Here, in this paper, we have proposed a simple tool to provide relative ranking for different image sequences. The purpose has been achieved by the ``CCUC'' method with easy computation and its availability.

\section{CONCLUSION}
This paper focuses on the problem faced by video monitoring system. Because of limited resources, the change levels of different image sequences should be compared. We formulate the change comparison problem and define how to determine the change level. Uncertainty Coefficient is used to obtain the position in one image sequence where the most abrupt is. Also it is applied to compare two image sequences to obtain the one with more abrupt change. The experiments have shown the CCUC method applicable and computationally efficient. Thus, CCUC is quite suitable to provide a ranking in video monitoring system.
\par
There are several points in our work can be improved further. Firstly, to our best knowledge, since most studies focus on the change detection instead of change comparison, there are few researches considering the change comparison between different image sequences. We cannot find enough methods to compare the power of our new method to. We show that our method is applicable, but we cannot say our new method is better without comparison. Secondly, in order to reduce computation complexity, we do not use complex methods such as neural network, hidden markov model. The use of more complex methods may improve the performance of comparison, but the computation may be not as simple as CUCC. Thirdly, we define the usage of Uncertainty Coefficient as U(previous one $|$ later one), the relative position of previous image and later image will somewhat influence the performance of comparison. Further, intuitively, $\operatorname{I}(previous;later)/\sqrt{\operatorname{H}(previous)\operatorname{H}(later)} $ may also improve the performance of comparison.

\section*{Acknowledgement}

This research was finished in Massachusetts Institute of Technology(MIT) under
the guidance of Prof. Berthold K.P. Horn and Dr. Yajun Fang. 
Thanks for the kindly help of Prof. Berthold K.P. Horn and Dr. Yajun Fang during my visit in Massachusetts Institute of Technology.
\par

\bibliographystyle{IEEEtran}  
\bibliography{uv}

\end{document}